\documentclass[10pt]{iopart}
\usepackage{iopams,braket}
\usepackage{graphicx,overpic,floatrow}
\usepackage[caption=false]{subfig}
\begin{document}

\title{Super- and subradiance of clock atoms in multimode optical waveguides}
 
\author{Laurin Ostermann}
\address{Institut f\"ur Theoretische Physik, Universit\"at Innsbruck\\Technikerstra{\ss}e 21, A-6020 Innsbruck, Austria}
\ead{Laurin.Ostermann@uibk.ac.at}

\author{Cl\'ement Meignant}
\address{Laboratoire d'Informatique de Paris 6, CNRS, Sorbonne Université, 4 place Jussieu, 75005 Paris, France}

\author{Claudiu Genes}
\address{Max Planck Institute for the Science of Light\\Staudtstra{\ss}e 2, D-91058 Erlangen, Germany}

\author{Helmut Ritsch}
\address{Institut f\"ur Theoretische Physik, Universit\"at Innsbruck\\Technikerstra{\ss}e 21, A-6020 Innsbruck, Austria}

\vspace{10pt}
\begin{indented}
\item[] \today
\end{indented}

\begin{abstract}
The transversely confined propagating modes of an optical fiber mediate virtually infinite range energy exchanges among atoms placed within their field, which adds to the inherent free space dipole-dipole coupling. Typically, the single atom free space decay rate largely surpasses the emission rate into the guided fiber modes. However, scaling up the atom number as well as the system size amounts to entering a collective emission regime, where fiber-induced  superradiant spontaneous emission dominates over free space decay. We numerically study this super- and subradiant decay of highly excited atomic states for one or several transverse fiber modes as present in hollow core fibers. As particular excitation scenarios we compare the decay of a totally inverted state to the case of  $\pi/2$ pulses applied transversely or along the fiber axis as in standard Ramsey or Rabi interferometry. While a mean field approach fails to correctly describe the initiation of superradiance, a second-order approximation accounting for pairwise atom-atom quantum correlations generally proves sufficient to reliably describe superradiance of ensembles from two to a few hundred particles. In contrast, a full account of subradiance requires the inclusion of all higher order quantum correlations. Considering multiple guided modes introduces a natural effective cut-off for the interaction range emerging from the dephasing of different fiber contributions.
\end{abstract}

\vspace{2pc}

\submitto{Focus on Nanoscale Quantum Optics}

\section{Introduction}
Over the last decades, high precision spectroscopy as one of the leading fields in quantum metrology~\cite{giovannetti2011advances} has brought forward ground-breaking results such as the world's best atomic clocks~\cite{ushijima2015cryogenic, ye2008quantum, liu2017realization} alongside all kinds of state-of-the-art devices as magnetometers, accelerometers or other quantum sensing devices~\cite{giovannetti2011advances}. To improve measurements one seeks to minimize quantum projection noise diminishing as $1/\sqrt{N}$ with an increasing number $N$ of emitters. However, when many atoms are confined in a small volume this leads to a compromise between reducing projection noise and introducing detrimental interaction effects owing to larger atom densities and the resulting increased dipole-dipole couplings. One close to ideal approach for clocks is to place the atoms into a magic wavelength optical lattice with one particle per site~\cite{campbell2017fermi}, where interactions occur via the remaining resonant dipole-dipole coupling only~\cite{kramer2016optimized}. This leads to unprecedented precision and accuracy at the cost of a high preparation effort.

The platform of optical fibers has attracted a lot of interest in the last years with tapered nano-fibers used for trapping and observation of single emitters from cold atoms~\cite{vetsch2010optical,tiecke2014nanophotonic} to nanoparticles ~\cite{liebermeister2014tapered} with applications as a single particle optical switch~\cite{o2013fiber} or an optical isolator~\cite{lodahl2017chiral}. Similarly, atoms have been trapped inside hollow core photonic crystal fibers for spectroscopy~\cite{vorrath2010efficient,okaba2014lamb,epple2014rydberg,xin2018atom} or non-linear optics~\cite{russell2014hollow} and even single molecules were coupled to dielectric nano- waveguides~\cite{tuerschmann2017chip}. In parallel, a great deal of theoretical investigations of the effective collective atom-field dynamics, the resulting light forces and possible applications in these systems have been carried out ranging from self organization~\cite{chang2013self, griesser2013light,holzmann2015collective, holzmann2016tailored}, to nonlinear scattering~\cite{domokos2002quantum, horak2003giant}. Applications of the collective radiative behavior into the light modes also point towards improved photonic memories~\cite{asenjo2017exponential}. Other recent studies include~\cite{tuerschmann2017chip,haakh2016polaritonic, faez2014coherent,goban2015superradiance}.

Along these lines, the idea to use optical fibers and in particular hollow core photonic crystal fibers~\cite{ilinova2017feasibility} for high precision spectroscopy was put forward by Okaba et.\ al.\ in 2014~\cite{okaba2014lamb}. This should allow for a collision-free setup of a large number of atoms trapped in a one dimensional optical lattice with a filling factor of $\leq 1$, which can be addressed in a uniform way via fiber guided modes. Here, confinement comes at the expense of some remaining atom-wall interactions and the sometimes unwanted enhanced long range dipole-dipole coupling. In their study Okaba et.\ al.\ saw clear indications of collective radiative emission as well as energy shifts and studied the effect of tight confinement and atomic interactions with the fiber wall. Nevertheless, even in their first pioneering study they were already able to reduce the relative uncertainty in measuring the $^1S_0 \to ^3P_1$ transition in $^{88}\mathrm{Sr}$ to the order of $10^{-13}$.

For long lived atomic transitions as used in optical clocks, free space spontaneous emission is only a small hinderance to precision measurements. However, at higher densities (leading to a larger optical depth) even in free space scenarios, collective radiative effects can be increased drastically~\cite{ostermann2012cascaded,ostermann2013protected,ostermann2014protected}. Such a fast collective decay (dubbed superradiance) can substantially decrease the sensitivity of a Ramsey type measurement. However, collective subradiant states coexist with the superradiant ones and the choice of favorable lattice and excitation geometries can lead to improved precision  via subradiant state protection~\cite{Bromley2016Collective,plankensteiner2015selective}.

In this paper we study superradiant/subradiant effects occurring in a 1D optical fiber geometries, where in addition to the collective coupling to the 3D free space vacuum modes, the guided discrete modes of the fiber have the potential to introduce an additional virtually infinite-range dipole-dipole coupling. This is particularly true in single transverse mode geometries~\cite{chang2013self}. Recently, closely related long range superradiance effects were observed in evanescent wave nano-fibers~\cite{solano2017super}. Moreover, we generalize the treatment to multimode fibers with a particular emphasis on photonic bandgap hollow core geometries as used in~\cite{okaba2014lamb}. We consider a regular 1D chain of clock atoms coupled to several modes of a multimode optical fiber including free space spontaneous emission with inherent dipole-dipole coupling. As a reference point towards new physics we first study the enhanced spontaneous emission for a single mode fiber using various numerical approaches from the full master equation down to a simple mean field approach. In a further step we generalize these ideas to several fiber modes. While for small systems sizes the master equation can be solved directly, we have to resort to enhanced mean-field methods for larger particle numbers.

\section{Model}
\begin{figure} \label{fig:model}
\centering
\includegraphics[width=0.5\columnwidth]{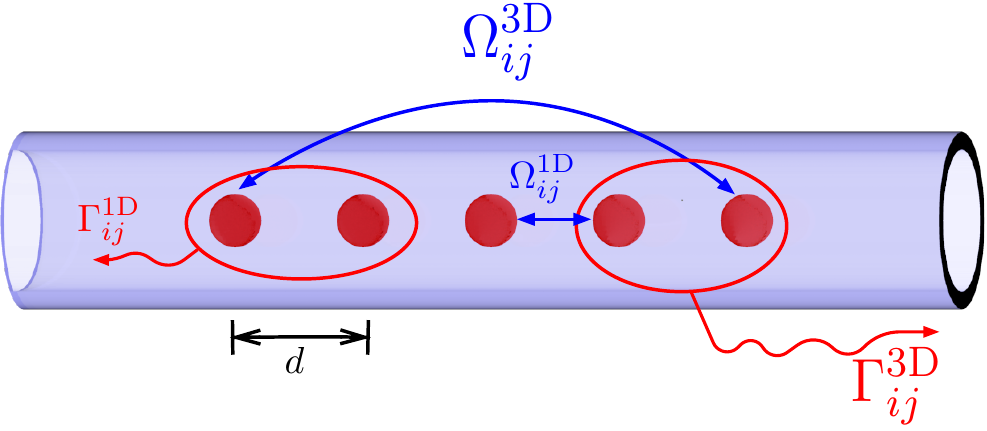}
\caption{{\it Model.} A collection of $N$ two-level emitters is positioned in a chain of lattice constant $d$ along the propagation direction of an optical fiber. Fiber mediated and free space coherent interactions are represented by mutual coupling strengths $\Omega^{1D}_{ij}$ and $\Omega^{3D}_{ij}$, while the collective spontaneous emission rates are denoted by $\Gamma^{1D}_{ij}$ and $\Gamma^{3D}_{ij}$.}
\end{figure}

We consider $N$ identical two-level emitters with transition frequency $\omega_0$ positioned at points $\lbrace \vec r_i \rbrace_{i=1}^N$ along a chain inside an (infinitely extended) optical fiber (see Fig.\ref{fig:model}) of cross-section $A=\pi a^2$ (where $a$ is an effective fiber radius). The atoms interact both with the free space modes as well as with the fiber guided modes. We will work in the approximation that the two processes can be separated which, after the elimination of the radiation field, results in dipole-dipole contributions stemming from the free-space modes ($\Omega^{3D}_{ij}$) and guided modes ($\Omega^{1D}_{ij}$). Additionally, incoherent mutual processes resulting from free space vacuum fluctuations are quantified by $\Gamma^{3D}_{ij}$ while the guided modes lead to mutual decay rates $\Gamma^{1D}_{ij}$.

\subsection{Effective master equation}
The full Hamiltonian for the system of atoms and radiation modes can be decomposed into contributions originating from coupling to 3D free space and 1D fiber modes and we have
\begin{equation}
H_\mathrm{full} = H_\mathrm{A} + H_\mathrm{F}^\mathrm{3D} + H_\mathrm{F}^\mathrm{1D} + H_\mathrm{int}^\mathrm{3D} + H_\mathrm{int}^{1D},
\end{equation}
with $H_\mathrm{A} = \omega_0 \sum_i \sigma^+_i \sigma^-_i$, where $\sigma^\pm_i$ denote the raising and lowering operators of the $i$-th emitter and we assume identical transition frequencies $\omega_0 = c k_0 = 2 \pi c / \lambda_0$ for all atoms. The free Hamiltonian of the radiation modes is split as $H_\mathrm{F}^\mathrm{3D} = \sum_{\vec k, \lambda} \omega_k a^\dagger_{\vec k, \lambda} a_{\vec k, \lambda}$ with the frequency $\omega = c \left| \vec k \right|$, the wave vector $\vec k$ and the polarization $\lambda$, and $H_\mathrm{F}^\mathrm{1D} = \sum_{\eta} \omega b^\dagger_\eta b_\eta$ where $\eta = \left( \omega, \lambda, \nu, f \right)$ with the frequency $\omega$, polarization $\lambda$, the mode index $\nu$ and the propagation direction $f = \pm 1$. We have assumed complete independence of the free-space and guided modes such that $\left[ a^\dagger_{\vec k, \lambda}, b_\eta \right] = \left[ a_{\vec k, \lambda}, b^\dagger_\eta \right] = 0$. Furthermore, we have $\left[ a_{\vec k, \lambda}, a^\dagger_{\vec k^\prime, \lambda^\prime} \right] = \delta_{\vec k \vec k^\prime} \delta_{\lambda \lambda^\prime}$ and $\left[ b_\eta, b^\dagger_{\eta^\prime} \right] = \delta_{\eta \eta^\prime}$. The two interaction terms in dipole and rotating wave approximation are given in the Appendix.

The elimination of the radiation modes is done in the standard quantum optics procedure~\cite{gardiner2004quantum} by formally integrating the equations of motion for $a_{\vec k, \lambda}$ and $b_\eta$ and inserting the solution into an integro-differential equation for an arbitrary operator of the emitters. Under the Markov approximation (see Ref.~\cite{ficek2002entangled,chang2012cavity}) one can show that the dynamics of the reduced system of $N$ atoms is properly described by the effective master equation
\begin{equation}
\dot \rho = i \left[ \rho, H \right] + \mathcal{L} [\rho],
\end{equation}
where the eliminated fields result in an effective dipole-dipole interaction via free space as well as via the guided modes, i.e.\
\begin{equation}
H = \omega_0 \sum_{i = 1}^N \sigma^+_i \sigma^-_i + \sum_{i \not = j} \left( \Omega_{ij}^\mathrm{3D} + \Omega_{ij}^{1D} \right) \sigma^+_i \sigma^-_j.
\end{equation}
The two contributions are fundamentally different. The free space coupling is finite range with considerable strengths only around the distances of the order of a wavelength,
\numparts
\begin{equation}
\Omega_{ij}^\mathrm{3D} \left( \xi = k_0 r_{ij} \right) = - \frac{3 \Gamma}{4} \left[ \frac{\cos \xi}{\xi} - \frac{\sin \xi}{\xi^2} + \frac{\cos \xi}{\xi^3} \right],
\end{equation}
while the guided modes yield a sum of infinite range interactions, periodic along the fiber's axis (as in~\cite{albrecht2018subradiant}),
\begin{equation}
\Omega_{ij}^\mathrm{1D} \left( \xi = k_0 z_{ij} \right) = \frac{\Gamma^\mathrm{1D}}{2} \ \sum_\nu \chi^\nu \sin \left( \beta^\nu \xi \right) \label{couplings1},
\end{equation}
\endnumparts
where $\chi^\nu$ expresses the coupling strength of each mode and $\beta^\nu$ is the propagation constant as defined in Eq.~(\ref{beta-eveq}) in the Appendix.

The coherent part is accompanied by incoherent processes responsible for dissipative dynamics such as super- and subradiance. We have 
\begin{equation}
\mathcal{L} \left[ \rho \right] = \frac{1}{2} \sum_{i, j} \left( \Gamma_{ij}^\mathrm{3D} + \Gamma_{ij}^{1D} \right) \left[ 2 \sigma^-_i \rho \sigma^+_j - \sigma^+_i \sigma^-_j \rho - \rho \sigma^+_i \sigma^-_j \right].
\end{equation}
As above, a finite range free-space contribution,
\numparts
\begin{equation}
\Gamma_{ij}^\mathrm{3D} \left( \xi = k_0 r_{ij} \right) = \frac{3 \Gamma}{2} \left[ \frac{\sin \xi}{\xi} + \frac{\cos \xi}{\xi^2} - \frac{\sin \xi}{\xi^3} \right],
\end{equation}
competes with the sum of infinite range fiber mediated decay rates,
\begin{equation}
\Gamma_{ij}^\mathrm{1D} \left( \xi = k_0 z_{ij} \right) = \Gamma^\mathrm{1D} \ \sum_\nu \chi^\nu \cos \left( \beta^\nu \xi \right) \label{couplings2}.
\end{equation}
\endnumparts
The single particle emission rates are
\begin{equation}
\Gamma = \frac{k_0^3 \mu^2}{3 \pi \epsilon_0}
\qquad
\Gamma^\mathrm{1D} = \frac{k_0 \mu^2}{\epsilon_0 A} = \alpha \Gamma,
\end{equation}
where $\alpha := \Gamma^\mathrm{1D}/\Gamma = 3 \pi / \left( k_0^2 A \right)$ and $\mu = \left| \vec \mu \right|$ is the atomic transition dipole moment, which we assume to be orthogonal to the propagation direction in the fiber, i.e.\ $\vec \mu \parallel \vec e_x$. The prefactor in the sum over all the guided modes of the 1D confined field reads
\begin{equation}
\chi^\nu = \left. \frac{\mathrm{d} \beta^\nu}{\mathrm{d k}} \right|_{k=k_0} \cdot \frac{A}{C^\nu} \left| E^\nu_x \right|^2
\end{equation}
and incorporates the field $E_x$ experienced by the dipoles, the normalization of the field $\int_0^\infty \mathrm{d}r \int_0^{2 \pi} \mathrm{d} \varphi \left| n E^\nu \right|^2 = C^\nu$ for each mode and the (scaled) group velocity. The field itself is obtained by solving Maxwell's equations in cylindrical coordinates for a cylindrical fiber of cross section $A = \pi a^2$ and refractive index $n_1 > 1$ cladded by vacuum $n_2 = 1$ (see appendix) in a similar manner to Ref.~\cite{le2014propagation}.

\subsection{Numerical methods}
The presented results mainly stem from numerical investigations based on the reduced master equation of the $N$ atoms under different initial conditions. As full simulations in the Hilbert space of $N$ two-level systems are computationally costly, we make use of different approximation methods factorizing in terms of individual operators or pairs of operator correlations. In increasing orders of computational time, the methods consist of: i) a mean field model (MF), ii) an augmented mean field model developed in Ref.~\cite{kramer2015generalized} including particle-particle correlations, which we dub 'mean field plus correlations' (MPC) and iii) an exact integration of the master equation (ME) as a basis of comparison~\cite{zoubi2010metastability}. For the exact master equation solution the number of equations that need to be integrated scales as $4^N$. The mean field approach reduces the effort in the integration to $3N$ equations, i.e.\ to the expectation values \ $\left \langle \sigma^x_j \right \rangle$, $\left \langle \sigma^y_j \right \rangle$ and $\left \langle \sigma^z_j \right \rangle$ for every atom ($j=1,...N$). In the MPC approach we include second order correlations and therefore our number of equations will scale as $3N \left( 2N+1 \right)$. More precisely, for MF we choose
\begin{equation}
\rho = \bigotimes_k \rho^{(k)},
\end{equation}
which effectively factorizes all correlations of the form $\left \langle A B \right \rangle \approx \left \langle A \right \rangle \left \langle B \right \rangle$, while the second order MPC approach is based upon the ansatz
\begin{equation}
\rho = \bigotimes_i \rho^{(i)} + \sum_{j < k} \left( \rho^{(j, k)} \otimes \bigotimes_{i \not = j, k} \rho^{(i)} \right)
\end{equation}
and we factorize $\left \langle A B C \right \rangle \approx -2 \left \langle A \right \rangle \left \langle B \right \rangle \left \langle C \right \rangle + \left \langle A \right \rangle \left \langle BC \right \rangle + \left \langle B \right \rangle \left \langle A C \right \rangle + \left \langle C \right \rangle \left \langle A B \right \rangle$. In order to investigate the validity of our approximations we compare the two inexact methods with the solution of the full master equation for reduced ensembles of not more than $10$ atoms. Where the dynamics are well contained within the approximations, we can perform numerical simulations up to a few hundred atoms (for MPC) or up to $10^4$ atoms (for MF).

\subsection{Initial states selected for the collective dynamics} \label{sec2.3}
All simulations describe the combined effects of guided and free space modes onto the collective decay of the atomic system. Interestingly, the dominant physical mechanisms not only depend on the geometric properties of the atoms and modes but they are very sensitive to the prepared initial state. We go well beyond the states of the single excitation manifold, which have been studied extensively in earlier work. In order to exhibit the most striking properties of superradiance, first, we consider the completely inverted state $\ket{eee \ldots e}$ and analyze its dynamics in the MF, MPC and ME treatments. In principle, this state can be prepared by a perfect $\pi$ pulse applied to all atoms simultaneously. Note, that all phase information of the excitation process is lost and we start with $\left \langle S_x \right \rangle = \left \langle S_y \right \rangle = 0$. We expect the appearance of a prototypical superradiant pulse with its maximal intensity scaling as $N^2$ after the initial time of building up coherence in the ensemble.

Yet, even if the excitation pulse was perfect for a single atom, interactions among the atoms during the preparation will prevent us from reaching inversion with unit fidelity. Thus, as a second case, we will consider an imperfect preparation where the system is initialized by a small non-vanishing transverse spin at $t=0$. This is equivalent to surpassing the initial difficulty of initializing a macroscopic coherence characteristic of the onset of superradiance. Consequently, that small contribution turns out to have a striking impact on the reliability of the approximation methods.

The third scenario we analyze is the case of a $\pi/2$ pulse as applied in the first step of a Ramsey sequence. Here, we prepare a state with a maximal transverse dipole moment as all dipoles are aligned in parallel in the transverse plane. This state is expected to exibit superradiant decay immediately. In this case the relative phase of the dipoles is crucial for the decay behaviour and we will study two important limiting cases, where the $\pi/2$ Ramsey pulses are applied either longitudinally or transversely with respect to the fiber axis. Transverse excitation prepares each atom in the same state $\left( \ket{g}_j+ \ket{e}_j \right) / \sqrt{2}$, while longitudinal driving will imprint progressing phases depending on the atomic positions, $\left( \ket{g}_j+ \exp \left( i \beta^0 k_0 r_j \right) \ket{e}_j \right) / \sqrt{2}$ and we chose the fundamental fiber mode for the driving.

\section{Single mode fiber: fully inverted state}
Geometries where a suitable choice of the inter-particle separation can lead to the cancelation of dipole-dipole shifts are of particular interest. We therefore focus on such a magic wavelength configuration with $d/\lambda_0 \approx 0.59$, as used in optical lattice experiments with trapped Strontium atoms driven on the $^1S_0$ to $^3P_0$ transition ~\cite{okaba2014lamb}. We then fix the effective fiber propagation constant to $\beta = 1.2$ and set the ratio $\alpha$ (largely defined by the fiber's diameter, which can be modified to tailor the relative role of the free-space and confined field collective interactions) to $0.75$. The dynamics is characterized by following the time evolution of the total excitation operator,
\begin{equation}
S^z = \sum_j \sigma^+_j \sigma^-_j,
\end{equation}
from the initial value of $N$.

\begin{figure} \label{fig:compare-full}
\centering
\sidesubfloat[]{\includegraphics[width=0.45\columnwidth]{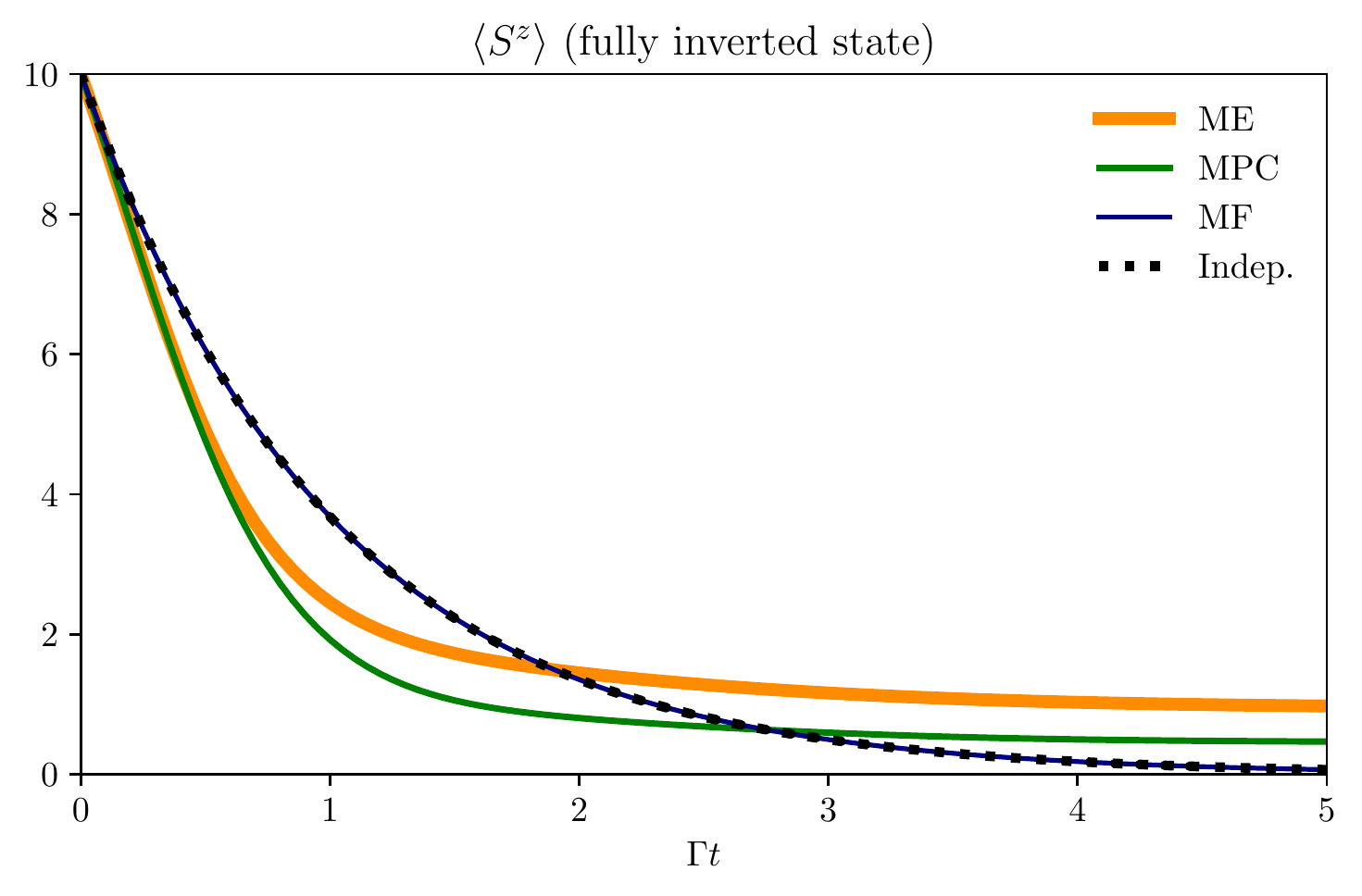}}
\hfill
\sidesubfloat[]{\includegraphics[width=0.45\columnwidth]{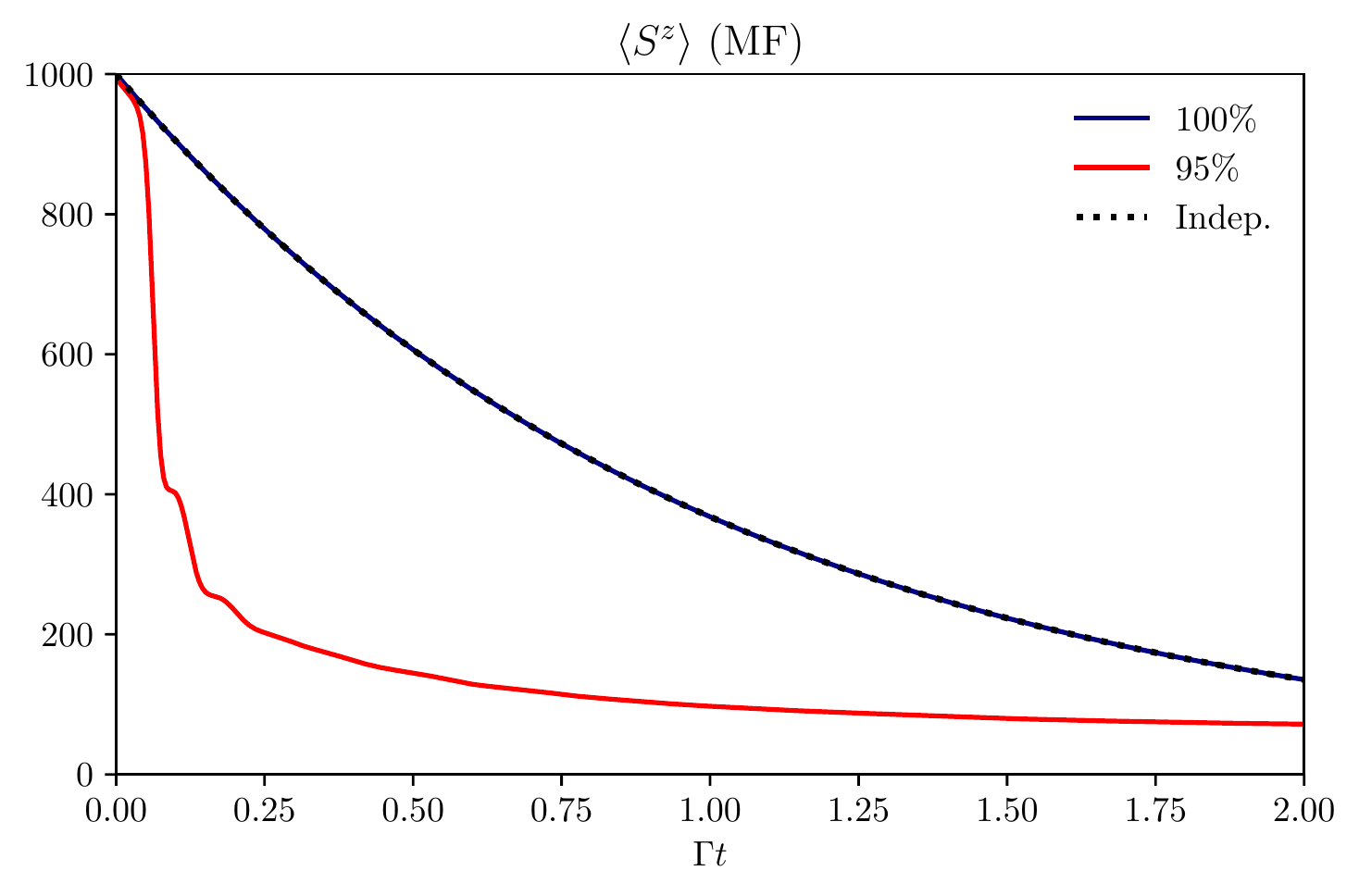}}
\caption{{\it Collective dissipative dynamics.} Collective decay of a regular chain of (a) ten atoms from the fully inverted state. The full master equation solution (ME) is compared to a mean field (MF) and a second order MPC solution as well as to the independent decay. The parameters are: $d/\lambda_0 = 0.59$, $\alpha = 0.75$, $\chi^1=1$ and $\beta^1 = 1.2$. (b) Mean field prediction of time evolution of 1000 atoms positioned at $d/\lambda_0 = 0.05$ from the fully inverted state $\ket{\psi} = \ket{e} \otimes \ldots \otimes \ket{e}$ in contrast to a state with only 95\% excitation of every individual emitter, i.e.\ $\ket{\psi_2} = \ket{\varphi} \otimes \ldots \otimes \ket{\varphi}$ with $\ket{\varphi} = \sin(\theta/2) \ket{g} + \cos(\theta/2) \ket{e}$, where $\theta = 0.95 \pi$. We see that assuming a tiny initial dipole moment suffices in order to capture the essence of superradiant decay via mean field calculations.}
\end{figure}

Generally, we find quantitative agreement between the exact ME solution and the MPC solution for short times and at least a qualitative one for longer times (see Fig.~\ref{fig:compare-full}a). However, while two particle correlations as included in the MPC method, are sufficient to capture the initial superradiant behavior, larger time subradiance implies more than two particle correlations. In contrast, the simple mean field result is often far off and usually only useful for independently decaying ensembles; this is because the mean field cannot capture collective effects such as entanglement or multipartite correlations. Especially when starting from a perfectly inverted state, the mean field approximation fails to correctly account for the buildup of correlations and coherence. However, assuming an initial macroscopic non-zero dipole (which is normally associated with the initial spontaneous build-up of macroscopic coherence in superradiance), the mean field approximation is closer to the real dynamics. We can achieve this regime by assuming imperfect excitation pulses close but not exactly at the $\pi$ value. This could be the result of non-negligible dipole-dipole interactions and collective decay during the excitation. 

Numerically we simulate the excitation under the following Hamiltonian (in an interaction picture removing the laser frequency) for the duration $\tau$ of the pulse
\begin{equation}
H_\mathrm{exc} = H + H_\mathrm{L} = \sum_{i\neq j} \Omega_{ij}\sigma_i^+\sigma_j^- + \Omega_\mathrm{R} \sum_i(\sigma_i^+ + \sigma_i^-),
\end{equation}
supplemented with the collective Linbdlad term describing decay. We have assumed resonant laser excitation with the Rabi frequency $\Omega_\mathrm{R}$. For an intended $\pi$ pulse we set the driving time $\tau$ by demanding that $\Omega_\mathrm{R} \tau=\pi/2$. The interatomic separation is fixed at the magic wavelength trapping distance $d / \lambda_0 = 0.59$. Note, that at this distance, the one dimensional collective decay (of infinite range) due to the fiber dominates over the free space collective decay (falling off as $(d/\lambda)^{-3}$). For the dynamics in Fig.~\ref{fig:compare-full}b, we consider a square pulse with Rabi frequency $\Omega_\mathrm{R} =100\Gamma^\mathrm{1D}$, for which the population of the excited state corresponds to a rotation with an angle $\theta = 0.95 \pi$ on the collective Bloch sphere. preserving some finite coherence in the spins as opposed to a perfect inversion. We compare independent decay for $1000$ spins prepared in the perfectly inverted state with the situation where the ensemble is prepared as described above via the application of a excitation laser. The perfect $\pi$ pulse leads to initial conditions $\langle \sigma_j^x \rangle(t=0)=0$ and $\langle \sigma_j^y \rangle(t=0)=0$ which evolve into $\langle \sigma_j^x \rangle (t) = \langle \sigma_j^y \rangle (t)= 0$ for all times. This implies that the atoms are decaying independently. In reality, spontaneous emission events lead to the emergence of a collective macroscopic dipole which speeds up the collective decay. As Fig.~\ref{fig:compare-full}b shows, this can be simulated by assuming the imperfect preparation scheme which creates an initial macroscopic dipole leading to faster decay than that of an independent ensemble.

\begin{figure}[t] \label{fig:ratio-full}
\centering
\sidesubfloat[]{\includegraphics[width=0.45\columnwidth]{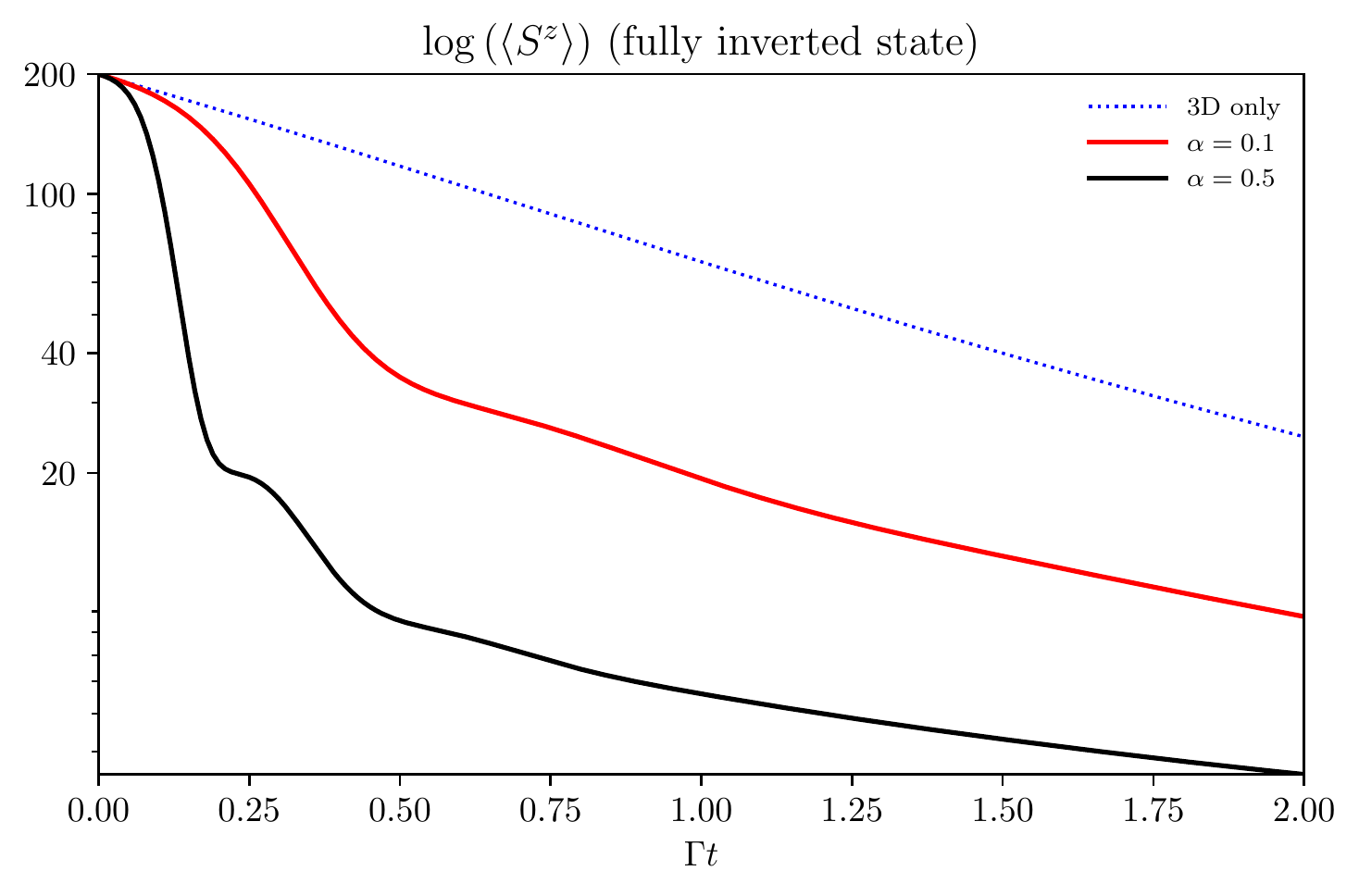}}
\hfill
\sidesubfloat[]{\includegraphics[width=0.45\columnwidth]{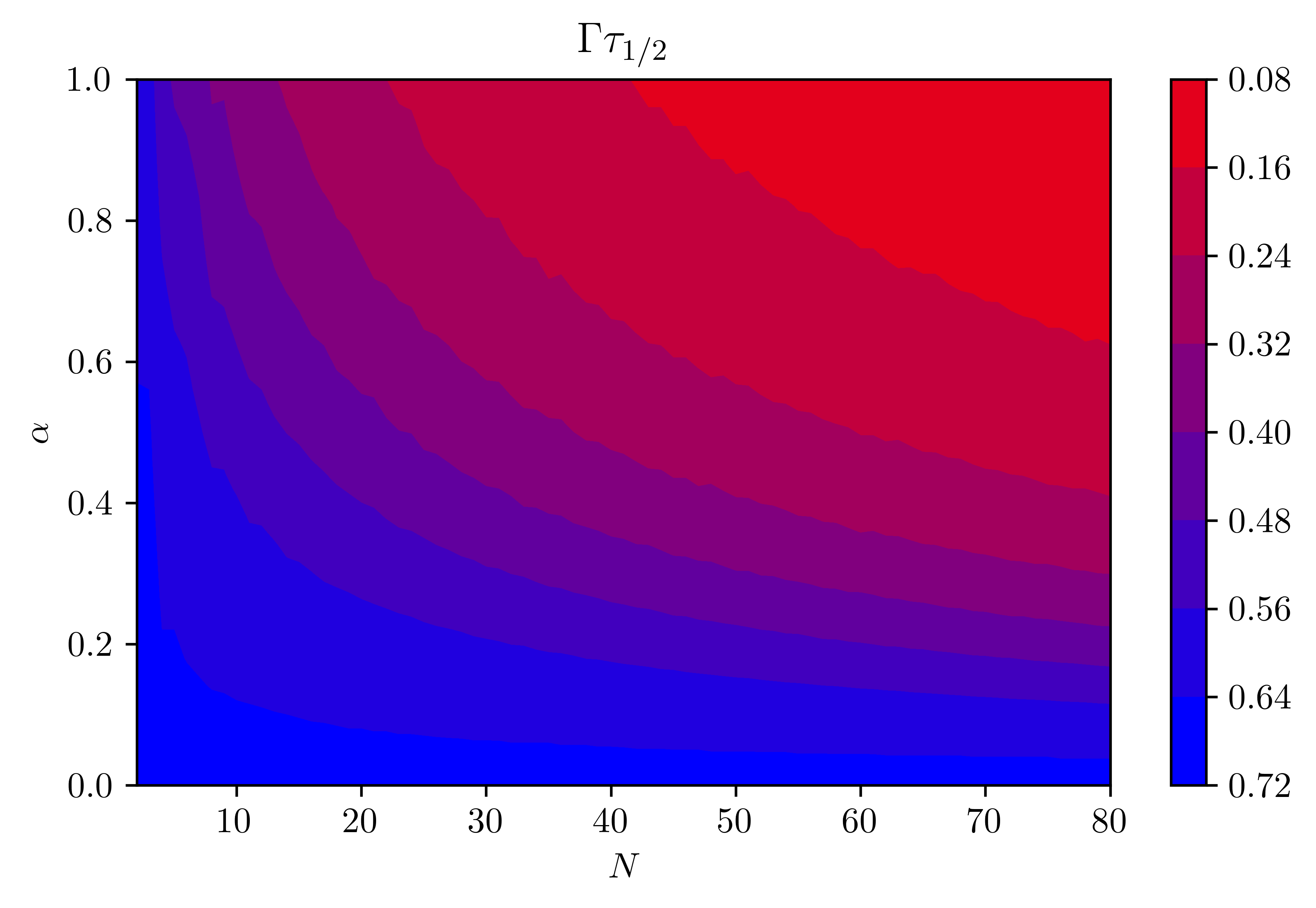}}
\caption{{\it Collective decay for different ratios $\alpha$ between 1D and 3D contributions and its scaling with atom numbers.} (a) Time evolution of the fully inverted state for the magic wavelength lattice interatomic distance of $d/\lambda_0 = 0.59$. MPC calculations show that the fiber-induced superradiance dominates with increasing 1D participation $\alpha$. (b) Time at which half the population has decayed out of the system starting at the fully inverted state as a function of the number of emitters $N$ and the ratio $\alpha$ in a chain of $d / \lambda_0= 0.59$. For an independently decaying ensemble this would be the case at $\Gamma t = 0.69$. For a fixed $\alpha$ (albeit small) increasing the number of atoms always leads to fiber-induced superradiance.}
\end{figure}
Let us now analyze the effect of the competition between free radiation modes induced decay versus collective decay aided by guided modes. For increasing values of the ratio $\alpha$, Fig.~\ref{fig:ratio-full} quantifies the additional superradiance induced by the fiber. The parameters chosen here are: $\lambda_0 = 689 \, \mathrm{nm}$, \ $\chi^1 = 1$ and the propagation constant is $\beta^1 = 1.2$. While the free space superradiance is practically negligible at magic wavelength separations, the long-range fiber mediated interactions, even for modest $\alpha$, can lead to considerable superradiant behavior. For example, for the parameters of Ref.~\cite{okaba2014lamb}, where $\lambda_0 = 689 \, \mathrm{nm}$ and $a = 20 \, \mu \mathrm{m}$, the resulting ratio $ \alpha \sim 10^{-4}$, requires a large number of atoms only, i.e.\ more than $10^4$, to show a sizable effect of the guided modes' superradiance. This aspect is more clearly analyzed in Fig.~\ref{fig:ratio-full}b with the direct conclusion that the more emitters are present in the ensemble, the more dominant the 1D decay is over the 3D free space decay. The figure shows the time at which half the population has decayed as a function of the number of emitters and the ratio between 1D and 3D decay when starting from the fully inverted state.

\section{Single mode fiber: half-inverted states}
\begin{figure}[t] \label{fig:compare-half}
\centering
\sidesubfloat[]{
\begin{overpic}[width=0.43\columnwidth]
	{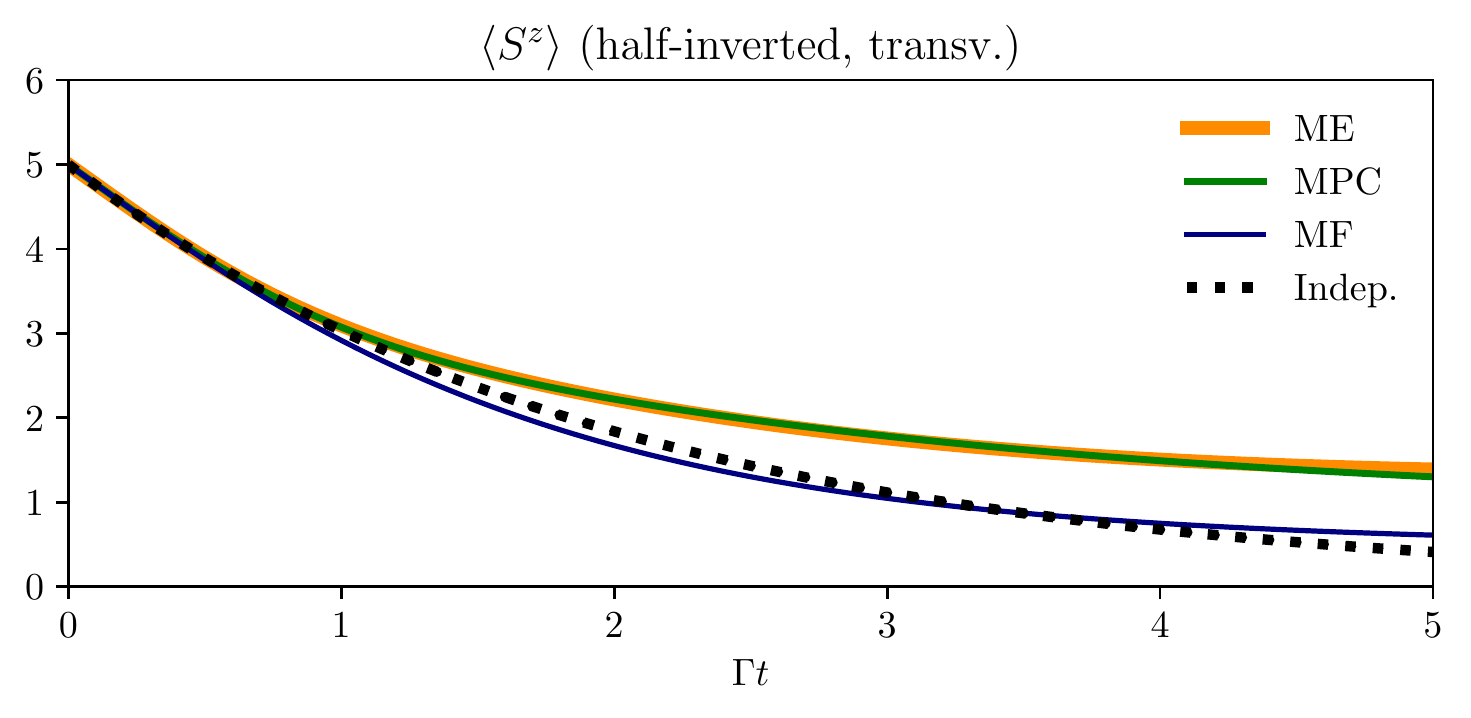}
	\put(44,31){\includegraphics[scale=0.15]{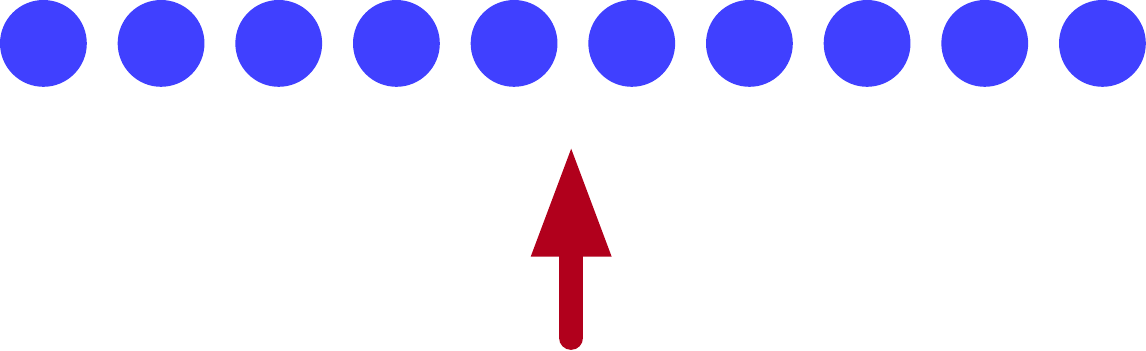}}
\end{overpic}
}
\hfill
\sidesubfloat[]{
\begin{overpic}[width=0.43\columnwidth]
	{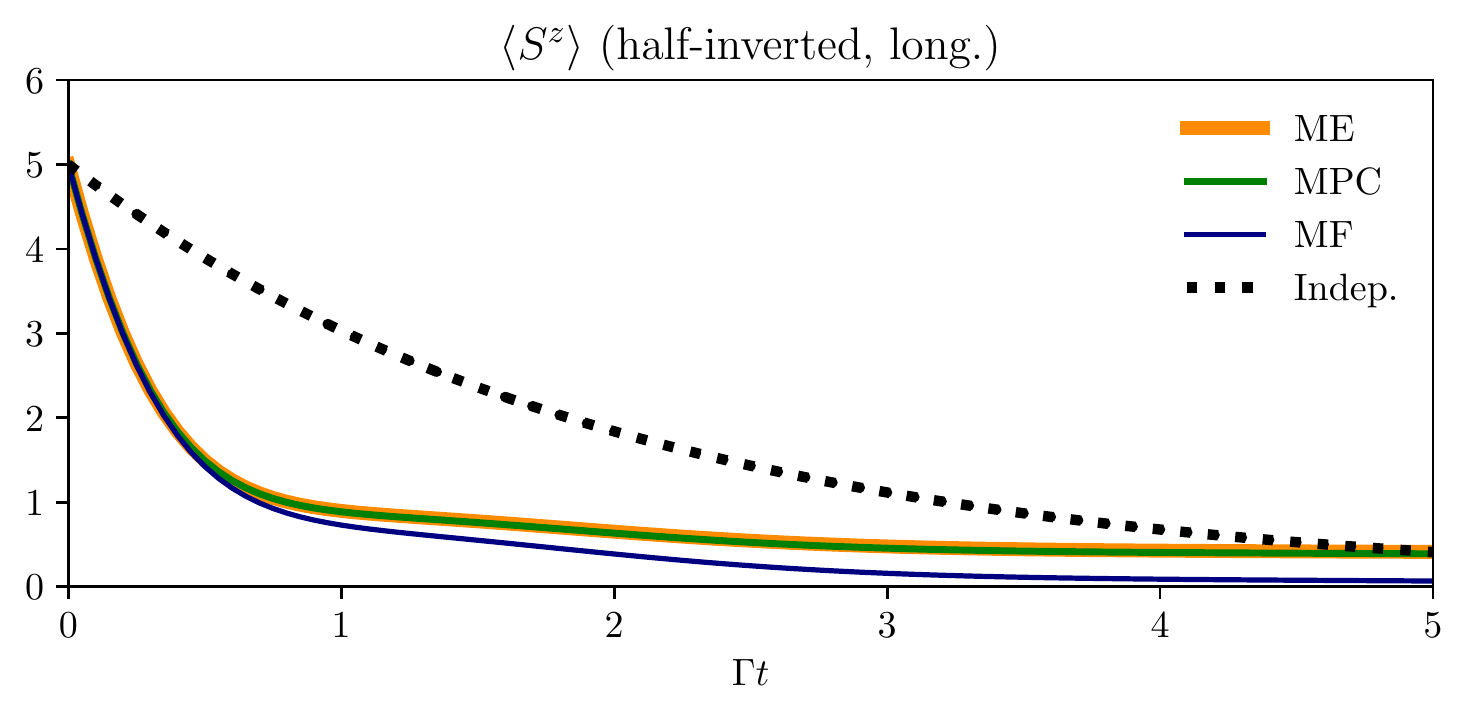}
	\put(35,38){\includegraphics[scale=0.15]{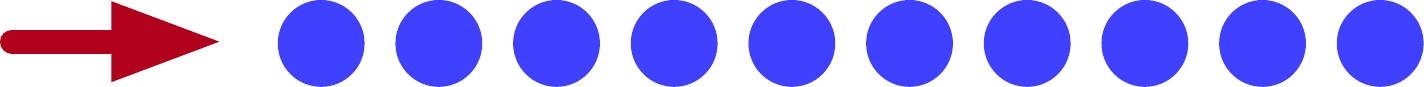}}
\end{overpic}
}
\caption{{ {\it Dynamics of half-inverted states.} Collective decay of a regular chain of ten atoms at $d/\lambda_0 = 0.59$ and $\alpha = 0.75$ from the half inverted state prepared by a (a) transversal and (b) longitudinal excitation pulse.} At the magic wavelength distance symmetric states are subradiant which leads to longer lifetimes for transverse excitation. In the longitudinal excitation case, the accumulated propagation phases counteract this behavior leading to superradiant decay, which limits the interrogation time in a Ramsey experiment.}
\end{figure}

In standard Ramsey interferometry, ensembles of atoms are prepared in states exhibiting maximal dipole moments along some direction, typically via the application of $\pi/2$ pulses. As described in Sec.~\ref{sec2.3}, at least two distinct procedures can be distinguished: i) transversal versus ii) longitudinal excitation. While transversal excitation imprints the same phase on all atoms, longitudinal excitation imprints different phases along the preferential direction defined by the fiber axis. We perform a detailed analysis of the numerical methods used for the two cases which we illustrate in Fig.~\ref{fig:compare-half}. For both situations, as concluded before, the MPC solution is a good estimate of the real dynamics while the MF solution depends strongly on the particularities of the initial state. In the transverse case, as shown in Fig.~\ref{fig:compare-half}, the dynamics are mostly subradiant. The effects largely stem from the fiber-induced collective decay.

Starting in the transversally pumped Ramsey state, which features a maximal dipole moment, the mean field solution is much closer to the exact result but will still not completely concur with the ME results on a long time scale. This initial state has a substantial overlap with states not radiating into the fiber which adds a long-lived component in the inversion that is again captured by the MPC corrections. In the other extremal case of a fiber excited Ramsey state, where we have an initial nonzero value for the Pauli operator in the $x$-direction all models capture superradiance along the fiber quite well, in contrast to the independent decay. Therefore, a mean field treatment represents a suitable choice to capture the system's dynamics when compared to the full master equation solution. We observe that in the experimentally most relevant case of a Ramsey sequence driven through the fiber, superradiance is strongly present limiting measurement time and thus precision.

\section{Collective decay in multimode fibers}
\begin{figure}[t] \label{fig:multimode}
\centering
\begin{overpic}[width=\columnwidth]
	{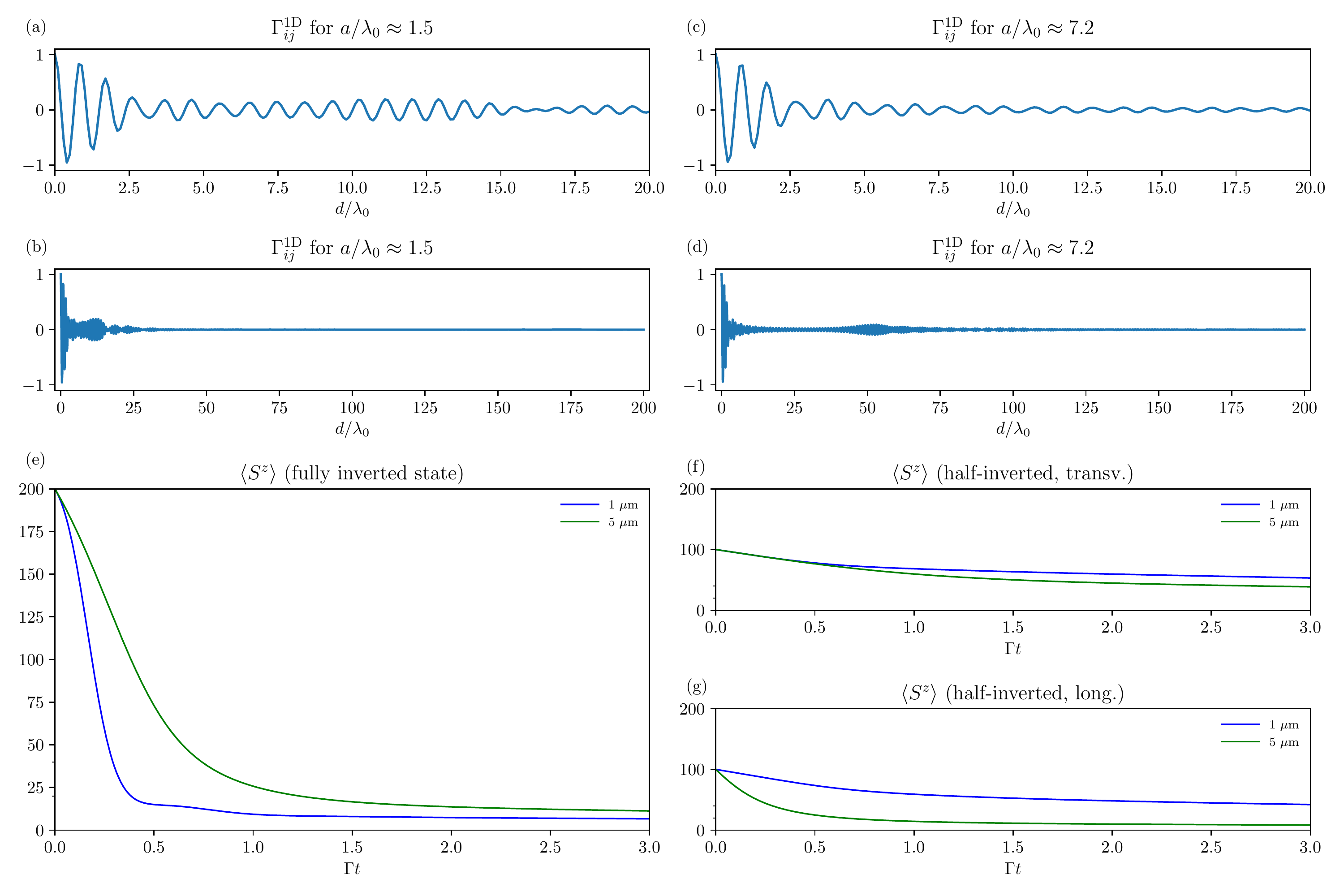}
	\put(80,26.5){\includegraphics[scale=0.10]{excitation-t.pdf}}
	\put(77.6,11.5){\includegraphics[scale=0.10]{excitation-l.pdf}}
\end{overpic}
\caption{{\it Collective radiative decay for multimode fibers.} The single mode infinite range interaction is modified by the weighted coupling to additional modes. We plot the distance dependence of the mutual decay rates of two atoms separated by $d$ for a small fiber diameter in (a) and (c), versus large diameter fibers in (b) and (d). For $\lambda_0 = 689 \, \mathrm{nm}$ the thin fiber of  $a = 1 \, \mu \mathrm{m}$ allows six guided modes, while a radius of $a = 5 \, \mu \mathrm{m}$ permits the propagation of 39 modes. The inclusion of additional modes leads to an effective decrease of the fiber-mediated atom-atom interaction range. In (e), (f) and (g) we show the time evolution of a fully (half) inverted system for those two fiber radii.}
\end{figure}
The diameter of optical fibers used in current experiments in conjunction with quantum emitters can range from $\sim0.5\mu$m to $\sim40\mu$m. While for sub-micron fibers the single mode approximation is well justified, for large diameter fibers one necessarily needs to consider more than one allowed mode. Of course, in the limit of very large diameters one would end up in the free space regime. This is equally true for evanescent fields of nanofibers as well as for the guided field inside a hollow core fiber. As the transverse mode functions are orthogonal, the Lindblad decay terms are independent and additional modes can be incorporated into our theory directly. Hence, in analogy to Eq.~(\ref{couplings1}) and Eq.~(\ref{couplings2}) the dipole-dipole interaction as well as the collective dissipation through the 1D guided modes is given by
\begin{displaymath}
\Omega_{ij}^\mathrm{1D} = \frac{\Gamma^\mathrm{1D}}{2} \ \sum_\nu \chi^\nu \sin \left( \beta^\nu \xi \right) \qquad \Gamma_{ij}^\mathrm{1D} = \Gamma^\mathrm{1D} \ \sum_\nu \chi^\nu \cos \left( \beta^\nu \xi \right),
\end{displaymath}
where $\nu$ is again the mode index and the propagation parameter $\beta^\nu$ and $\chi^\nu$ are particular to every mode and have to be determined numerically as a function of the fiber properties (see Appendix).

For simplicity, we choose a model of an infinitely extended cylindrical fiber where analytical expressions exist; in principle, one can use any suitable model calculation (e.g.\ using programs as COMSOL) in order to obtain the field in arbitrary optical structures, such as hollow core photonic crystals, tapered waveguides or nano fibers as done in~\cite{konorov2003waveguide}. The only thing that will change with respect to our formalism are the values of $\chi^\nu$ and $\beta^\nu$. As the collective couplings are a superposition of trigonometric functions of different periodicity since $\beta^\nu$ is different for each mode, by using different fiber diameters it is thus possible to tailor the interaction range. This is illustrated in Fig.~\ref{fig:multimode}, where we observe regions of more and less pronounced collective couplings. Consequently, instead of treating the multimode problem as a detrimental effect, we propose here that one can carefully include a number of modes to design interactions of tailorable range. The quite pronounced exponential decay of the couplings with distance can be achieved by allowing each mode to have a certain (Lorentzian) bandwith within the interval of allowed propagation constants as opposed to only one discrete contribution . The bandwidths are chosen in such a way that there is an overlap between the modes of a few percent~\cite{holzmann2016tailored}.

\section{Conclusions and Outlook}
For a reliable prediction of collective dynamics of quantum emitters periodically arranged within the field of a fiber, one has to include infinite range interactions via the effective 1D geometry of the system as well as finite range contributions mediated by the free 3D electromagnetic vacuum. In principle, no matter how small the atom-fiber coupling is, the 1D contribution will always dominate and introduce superradiant decay for large enough atom numbers. The combination of both effects introduces extra line shifts, which are small but important for clock applications. Interestingly, shifts and collective decay are reduced in multimode fibers, where only atoms within a finite range will participate so that divergences of shifts and decay rates in the limit of large particle numbers are prevented. In some regimes where the build-up of particle-particle correlations can be neglected, mean field calculations can give reasonable quantitative estimates of the collective system dynamics. However, two-particle correlations can quite reliably be accounted for in the MPC method derived in Ref.~\cite{kramer2015generalized} and allow for a general simulation of the correct dynamics for hundreds of particles. Yet, some aspects of subradiance seem to necessarily include more than particle-particle quantum correlations falling outside the validity of the MPC approach. As a bottom line, a hollow core multimode fiber system based on a fiber guided magic lattice trap could thus be a favourable way for a technically simplified but still accurate and precise implementation of an optical clock. It requires, however, a careful design of the excitation process and lattice geometry to keep collective shifts and superadiance limited. In this respect a transformation to an active reference oscillator system looks very promising, if a clever way of pumping and collective cooling~\cite{xu2016supercooling} via higher order fiber modes could be engineered.

\section*{Acknowledgements}
Calculations were performed using the Quantum optics and Collective Spins Julia packages developed by Sebastian Kr\"{a}mer~\cite{kramer2018quantumoptics}. We thank M.~Moreno-Cardoner, A.~Asenjo-Garcia and D. Plankensteiner for insightful discussions. We acknowledge support by the Austrian Science Fund via project P29318-N27 (L.~O.) and the SFB FoQus project F4013 (H.~R.). C.~G. thanks the University of Innsbruck for hospitality and the Max Planck Society for financial support.

\section*{References}
\providecommand{\newblock}{}

\appendix
\section{Fiber field and interactions}
Our system consists of a cylindrical infinitely long vacuum clad fiber of radius $a$ and refractive index $n_1$ embedded in vacuum ($n_2 = 1$) with $N$ two-level atoms ($\omega_0$, $\mu$) fixed at positions $\lbrace z_j \rbrace_{j=1}^N$ along the fiber. 

\subsection{Electric field in and outside the fiber}
For the field we solve Maxwell's equations, which we sketch briefly. For the electric field along the fiber axis we have the wave equation in radial coordinates with the propagation parameter $\beta$
\begin{equation}
\Box E_z = 0.
\end{equation}
With the ansatz
\begin{equation}
E_z = A f(r) e^{i \nu \varphi} \cdot e^{i \left( \omega t - \beta z \right)}
\end{equation}
this reduces to
\begin{equation}
 \left( \partial^2_r + \frac{1}{r} \partial_r + \frac{1}{r^2} \partial^2_\varphi + \left( n^2 k^2 - \beta^2 \right) \right) E_z = 0
\end{equation}
with the refractive index $n$ ($n_1$, $n_2$) and $k^2 = \omega^2/c^2 = \epsilon_0 \mu_0 \omega^2$. The same equation holds for the magnetic field $H_z$ and from that we can calculate the other two electric field components as
\begin{eqnarray}
E_r =& \frac{-i}{\kappa^2} \left( \beta \, \partial_r E_z + \frac{\omega \mu_0}{r} \, \partial_\varphi H_z \right), \\
E_\varphi =& \frac{-i}{\kappa^2} \left( \frac{\beta}{r} \, \partial_\varphi E_z - \omega \mu_0 \, \partial_r H_z \right).
\end{eqnarray}

For the radial component $f(r)$ we obtain the differential equation for the Bessel functions
\begin{equation}
\partial_r^2 f + \frac{1}{r} \partial_r f + \left( \kappa^2 - \frac{\nu^2}{r^2} \right) f = 0
\end{equation}
This results in the fields
\begin{equation}
f(r) = \cases{A J_\nu (hr) & $0 < r < a$ \\ C K_\nu (qr) & $a < r < \infty$},
\end{equation}
where $h = \sqrt{n_1^2k^2 - \beta^2}$ and $q = \sqrt{\beta^2 - n_2^2 k^2}$. $J_\nu$ denotes the Bessel function of the first kind, while $K_\nu$ is the modified Bessel function of the second kind. Similar solutions exist for $H_z$ (constants $B$ and $D$). Thus, we can construct the electric fields for $r < a$
\begin{eqnarray}
E_z&=& A J_\nu (hr) e^{i \nu \varphi} \\
E_r &=& \frac{-i}{h^2} \left( \beta h A J^\prime_\nu (hr) + i \omega \mu_0 \frac{\nu}{r} B J_\nu (hr) \right) e^{i \nu \varphi} \\
E_\varphi &=& \frac{-i}{h^2} \left( i \beta \frac{\nu}{r} A J_\nu (hr) - h \omega \mu_0 B J^\prime_\nu (hr) \right) e^{i \nu \varphi}
\end{eqnarray}
and for $r > a$
\begin{eqnarray}
E_z&=& C K_\nu (qr) e^{i \nu \varphi} \\
E_r &=& \frac{-i}{q^2} \left( \beta q C K^\prime_\nu (qr) + i \omega \mu_0 \frac{\nu}{r} D K_\nu (qr) \right) e^{i \nu \varphi} \\
E_\varphi &=& \frac{-i}{q^2} \left( i \beta \frac{\nu}{r}C K_\nu (qr) - q \omega \mu_0 D K^\prime_\nu (qr) \right) e^{i \nu \varphi}.
\end{eqnarray}
From the boundary condition for the tangential field components at $r=a$ we find
\begin{eqnarray}
C = \frac{J_\nu (ha)}{K_\nu (qa)} A \qquad D = \frac{J_\nu (ha)}{K_\nu (qa)} B \\
B = \frac{i}{\nu} \frac{a h q \left( n_1^2 q J^\prime_\nu (ha) K_\nu (qa) + i n_2^2 h J_\nu (ha) K^\prime_\nu (qa) \right)}{\omega (n_1^2 - n_2^2) \mu_0 \beta J_\nu (ha) K_\nu (qa)} A.
\end{eqnarray}

Furthermore we obtain an eigenvalue equation for the propagation parameter $\beta$ from the condition that the linear system for $A$, $B$, $C$, $D$ has to have a solution, i.e.
\begin{eqnarray} \label{beta-eveq}
\left( \frac{J^\prime_\nu (ha)}{ha J_\nu(ha)} + \frac{K^\prime_\nu (qa)}{qa K_\nu (qa)} \right)
\left( \frac{n_1^2 J^\prime_\nu (ha)}{ha J_\nu(ha)} + \frac{n_2^2 K^\prime_\nu (qa)}{qa K_\nu (qa)} \right) \\
-\nu^2 \left( \frac{1}{(ha)^2} + \frac{1}{(qa)^2} \right)^2 \frac{\beta^2}{k^2} = 0. \nonumber
\end{eqnarray}

This leaves us with one free parameter $A$ which we fix by the normalization condition
\begin{equation}
\int_0^{2 \pi} \mathrm{d} \varphi \int_0^\infty r \, \mathrm{d}r \, n^2 \left| E \right|^2 = 1,
\end{equation}
where $n = n_1$ for $r < a$ and $n= n_2$ for $r \geq a$.

\subsection{Interaction with atoms}
We will now derive a master equation for the atoms where coherent and incoherent atom-atom interactions stem from the collective coupling to both free and guided radiation modes. We start from the Hamiltonian
\begin{equation}
H_\mathrm{full} = H_\mathrm{A} + H_\mathrm{F}^\mathrm{3D} + H_\mathrm{F}^\mathrm{1D} + H_\mathrm{int}^\mathrm{3D} + H_\mathrm{int}^{1D},
\end{equation}
with $H_\mathrm{A} = \omega_0 \sum_i \sigma^+_i \sigma^-_i$, where $\sigma^\pm_i$ denote the raising and lowering operators of the $i$-th emitter and
\begin{equation}
H_\mathrm{F}^\mathrm{3D} = \sum_{\vec k, \lambda} \omega_k a^\dagger_{\vec k, \lambda} a_{\vec k, \lambda} \\
H_\mathrm{F}^\mathrm{1D} = \sum_{\eta} \omega b^\dagger_\eta b_\eta
\end{equation}
where $\eta = \left( \omega, \lambda, \nu, f \right)$ with the frequency $\omega$, polarization $\lambda$, the mode index $\nu$ and the propagation direction $f = \pm 1$. We have assumed complete independence of the free-space and guided modes such that $\left[ a_{\vec k, \lambda}^\dagger, b_\eta \right] = \left[ a_{\vec k, \lambda}, b_\eta^\dagger \right] = 0$. The two interaction terms in dipole and rotating wave approximation read
\begin{eqnarray}
\fl H_\mathrm{int}^{3D} = - \sum_{i = 1}^N \vec d_i \cdot \vec E^\mathrm{3D} ( \vec r_i) \approx - i \sum_{i=1}^N \sum_{\vec k, \lambda} g_{\vec k \lambda} \ \vec e_{\vec k, \lambda} \cdot \vec \mu \ \left( a_{\vec k, \lambda} e^{i \vec k \vec r_i} \sigma^+_i - h.c. \right) \\
\fl H_\mathrm{int}^{1D} \approx - i \sum_{i=1}^N \int_0^\infty \sum_\eta g_\eta \ \vec \mu \cdot \left( \vec e_\eta b_\eta e^{i f \beta^\nu k z_i} \sigma^+_j - \mathrm{h.c.} \right),
\end{eqnarray}
where $g_{\vec k \lambda} = \sqrt{\omega_k/2 \epsilon_0 V}$ and
\begin{equation}
g_\eta = \sqrt{\frac{\omega}{2 \epsilon_0 v_g}} \left[ \vec \mu \cdot E (r_j, \varphi_j) \right]
\end{equation}
where $v_g = \mathrm{d} \omega / \mathrm{d} \beta$ is the group velocity in accordance with~\cite{le2014propagation}. As the 3D and the 1D parts are linearly superimposed and thus can be treated independently, we will now focus on the 1D part only whereas the 3D derivation can be found e.g.\ in~\cite{ficek2002entangled}. Furthermore we will restricgt ourselves to one mode, i.e.\ neglect the sum over $\nu$ for now. From this we can calculate the time evolution of our field mode by solving
\begin{equation}
\dot b_\eta= i \left[ H, b_\eta \right] = -i \omega b_\eta + \frac{1}{\sqrt{2 \pi}} \sum_{j=1}^N g^*_\eta \ e^{-i \beta (\omega) z_j} \ \sigma^-_j,
\end{equation}
which yields
\begin{equation}
b_\eta (t) = b_\eta(0) \ e^{-i \omega t} + \frac{1}{\sqrt{2 \pi}} \int_0^t \mathrm{d}t^\prime \ e^{-i \omega ( t - t^\prime )} \, g^*_\eta \ \sum_{j=1}^N e^{-i \beta (\omega) z_j} \sigma^-_j \left( t^\prime \right).
\end{equation}

We continue by writing down the equation of motion for an arbitrary system operator, i.e.\ any operator acting on our ensemble of two-level emitters. We substitute for $b_\eta$, replace $\sum_\eta \to \int \mathrm{d} \omega \sum_{\lambda, \nu, f}$ and have
\begin{eqnarray}
\dot O =& i \left[ H_\mathrm{A}, O \right] + \frac{1}{\sqrt{2 \pi}} \int_0^\infty \mathrm{d} \omega \sum_{\lambda, f} \sum_j \\
 & \left( g_\eta b_\eta e^{i f \beta (\omega) z_j} \left[ \sigma^+_j, O \right] \right. - \left. g^*_\eta b^\dagger_\eta e^{- i f \beta (\omega) z_j} \left[ \sigma^-_j, O \right] \right). \nonumber
\label{O-timev}
\end{eqnarray}
Let us now focus on the interaction part and replace $b_\eta(t)$ and $b^\dagger_\eta$ with the inhomogeneous part from the expression from above (the homogeneous part is the in-field and we'll neglect it too)
\begin{eqnarray}
\dot O =& \frac{1}{2 \pi} \int_0^\infty \mathrm{d} \omega \sum_{\lambda, f} \sum_{j, k}  \\
 & \left( \int_0^t \mathrm{d} t^\prime e^{-i \omega (t-t^\prime)} \left| g_\eta \right|^2 e^{if \beta (\omega) (z_j - z_k)} \sigma^-_k (t^\prime) \left[ \sigma^+_j, O \right] \right. \nonumber \\
- & \left. \int_0^t \mathrm{d} t^\prime e^{i \omega (t-t^\prime)} \left| g_\eta \right|^2 e^{- i f \beta (\omega) (z_j - z_k)} \left[ \sigma^-_j, O \right] \sigma^+_k (t^\prime) \right) \nonumber
\end{eqnarray}
At this point we perform the Markov approximation (similar for $\sigma^-_j$)
\begin{equation}
\sigma^+_j \left( t^\prime \right) \to e^{- \omega_0 \left( t - t^\prime \right) } \sigma^+_j (t)
\end{equation}
and rewrite the expression
\begin{eqnarray}
\dot O =& \frac{1}{2 \pi} \int_0^\infty \mathrm{d} \omega \sum_{\lambda, f} \sum_{j, k} \\
& \left( \int_0^t \mathrm{d} t^\prime e^{-i (\omega - \omega_0) (t-t^\prime)} \left| g_\eta \right|^2 e^{if \beta (\omega) (z_j - z_k)} \sigma^-_k \left[ \sigma^+_j, O \right] \right. \nonumber \\
- & \left. \int_0^t \mathrm{d} t^\prime e^{i (\omega - \omega_0) (t-t^\prime)} \left| g_\eta \right|^2 e^{- i f \beta (\omega) (z_j - z_k)} \left[ \sigma^-_j, O \right] \sigma^+_k \right) \nonumber
\end{eqnarray}

For the integral over $t^\prime$ we use the Sokhotski - Plemelj theorem as
\begin{equation}
 \int_0^t \mathrm{d} t^\prime e^{\mp i (\omega - \omega_0) (t-t^\prime)} = \mp i \mathcal{P} \frac{1}{\omega -\omega_0} + \pi \delta \left( \omega - \omega_0 \right)
\end{equation}

Let us pull out $\sum_{j,k}$ perform the sum over the propagation direction $f$, set $\vec \mu = \mu \cdot e_x$ (which kills the sum over $\lambda$) and proceed
\begin{eqnarray}
\dot O =& \sum_{j, k} \frac{1}{\pi} \int_0^\infty \mathrm{d} \omega \\
& \left( \left( - i \mathcal{P} \frac{1}{\omega -\omega_0} + \pi \delta \left( \omega - \omega_0 \right) \right) \left| g_{\omega} \right|^2 \cos \left( \beta z_{jk} \right) \sigma^-_k \left[ \sigma^+_j, O \right] \right. \nonumber \\
- & \left. \left( i \mathcal{P} \frac{1}{\omega -\omega_0} + \pi \delta \left( \omega - \omega_0 \right) \right) \left| g_{\omega} \right|^2 \cos \left( \beta z_{jk} \right) \left[ \sigma^-_j, O \right] \sigma^+_k \right) \nonumber
\end{eqnarray}

In this form we can roll the equation back to the time evolution of the density operator and obtain a Lindblad-type master equation reading
\begin{equation}
\dot \rho = i \left[ \rho, \sum_{j \not = k} \Omega^\mathrm{1D}_{jjk} \sigma^+_j \sigma^-_k \right] + \sum_{jk} \frac{\Gamma^\mathrm{1D}_{jk}}{2} \left( 2 \sigma^-_j \rho \sigma^+_k - \sigma^+_j \sigma^-_k \rho - \rho \sigma^+_j \sigma^-_k \right),
\end{equation}
where we have neglected the infinite Lamb shift for $j = k$. The two coupling parameters are calculated as
\begin{equation}
\frac{\Gamma^\mathrm{1D}_{jk}}{2} = \frac{1}{\pi} \int_0^\infty \mathrm{d} \omega \pi \delta \left( \omega - \omega_0 \right) \left| g_\omega \right|^2 \cos \left( \beta z_{jk} \right) = \frac{\omega_0 \mu^2 \left[ E_x \right|^2}{2 \epsilon_0 v_g} \cos \left( \beta z_{jk} \right)
\end{equation}
and for the dipole-dipole shift we employ the residue theorem, create a pole at $\omega_0 + i \epsilon$ with $\epsilon \to 0$, close the integration path in the uper half plane and take only the real part of the integral, such that
\begin{eqnarray}
\Omega^\mathrm{1D}_{jk} =& - \lim_{\epsilon \to 0} \Re \frac{1}{\pi} \oint \mathrm{d} \omega \frac{\left| g_\omega \right|^2}{\omega- \left( \omega_0 + i \epsilon \right)} \frac{1}{2} \left( e^{i \beta z_{jk}} + e^{-i \beta z_{jk}} \right) \\
=& - \frac{1}{2 \pi} \Re \, 2 \pi i \left| g_{\omega_0} \right|^2 e^{i \beta z_{jk}} \nonumber \\
=& \frac{\omega_0 \mu^2 \left[ E_x \right|^2}{2 \epsilon_0 v_g} \sin \left( \beta z_{jk} \right). \
\end{eqnarray}
At this point it is important to realize, that all functional dependencies, i.e.\ $E_x$, $v_g$ and $\beta$ are evaluated at the transition frequency $\omega_0$ and thus become a mere number.

\begin{figure} \label{fig:couplings}
\centering
\includegraphics[width=0.75\columnwidth]{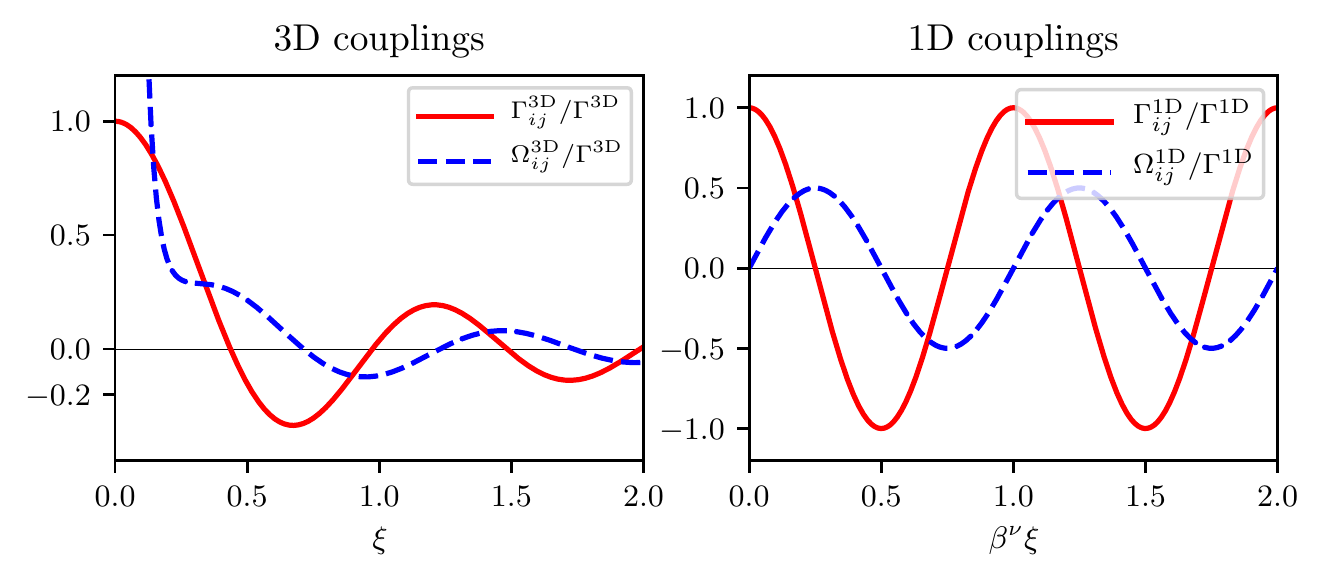}
\caption{{\it Collective couplings in 3D and 1D.} Spatial dependence of the 3D (left) and 1D (right) coherent/incoherent coupling strengths as a function of distance between atoms as a function of the normalized distance $\xi = k_0 d$. Observe, that the 3D free space couplings decay with increasing distance, while the 1D contributions, shown for a single mode here, are periodic and of infinite range. Including several modes, finite ranges can be engineered as discussed below.}
\end{figure}

The couplings as a function of the interatomic distance $\xi = k_0 d$ are depicted in Fig.~\ref{fig:couplings}. For the 1D coupling we show the spatial behavior of one mode. Taking several modes into account leads to a superposition of trigonometric functions with different periodicities and amplitudes as is discussed in the paper.

\subsection{Multiple Modes}
In order to allow for multiple modes in our model, we modify the interaction Hamiltonian to accommodate for several modes, i.e.
\begin{equation}
H_\mathrm{int} = - \frac{i}{\sqrt{2 \pi}} \int_0^\infty \mathrm{d} \omega \sum_{f, \lambda, j} \sum_\nu \left( g_\eta b_\eta e^{i f \beta^\nu (\omega) z_j} \sigma^+_j - \mathrm{h.c.} \right).
\end{equation}

Doing this with the photonic operators for different modes commuting we get slightly modified coupling constants in a very similar derivation, which read
\begin{eqnarray}
\Gamma_{jk} =& \sum_{\nu} \frac{\omega_0 \mu^2 \left[ E^\nu_x \right|^2}{\epsilon_0 v^\nu_g} \cos \left( \beta^\nu z_{jk} \right) \\
\Omega_{jk} =& \sum_{\nu} \frac{\omega_0 \mu^2 \left[ E^\nu_x \right|^2}{2 \epsilon_0 v^\nu_g} \sin \left( \beta^\nu z_{jk} \right)
\end{eqnarray}

\begin{figure} \label{fig:dispersion}
\centering
\includegraphics[width=0.7\columnwidth]{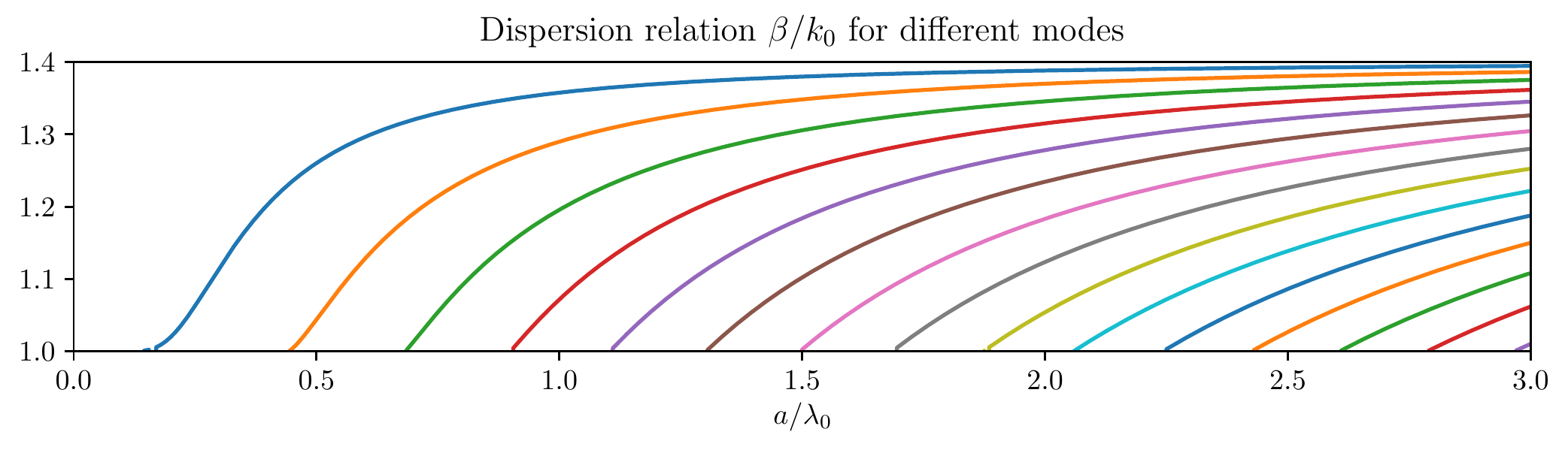}
\caption{{\it Dispersion relations.} Propagation constant $\beta^\nu/k_0$ for different ratios of fiber radius vs. wavelength of the propagating light $a/\lambda_0$. For a given fiber radius and a chosen atomic transition, a vertical cut of this plot gives all the propagation constants corresponding to the different allowed modes. The narrower the fiber is in relation to the wavelength of the atomic transition targeted, the less modes can propagate. The slope of these curves will give the scaled group velocity that appears as a factor within $\chi^\nu$.}
\end{figure}

For an impression of how the different modes will propagate, Fig.~\ref{fig:dispersion} depicts the dispersion relation as a function of the ratio between the fiber's radius and the wavelength of the propagating light. In a thin nano fiber, where $a/\lambda_0 < 0.5$, only a few modes will be allowed, up to the extremal case, where only the fundamental mode is permitted. The wider the fiber becomes in relation to the light's wavelength, the more modes will be able to propagate. For $a/\lambda_0 \approx 3$ we observe about 15 different modes. This implies that the tailoring of the interaction cannot only be done by placing the atoms at suitable distances to each other, but also by choosing a desired fiber radius.

\subsection{Numerical Parameters}
We now define a one-dimensional decay rate independent of the particular mode or the particular position of the atom as a prefactor
\begin{equation}
\Gamma_\mathrm{1D} = \frac{\omega_0 \mu^2}{\epsilon_0 c A},
\end{equation}
where $A$ is the cross section of the fiber $A= a^2 \pi$ and is compensated for by the normalization of the electric field. We can then write
\begin{equation}
\Gamma_{jk}^\mathrm{1D} = \Gamma_\mathrm{1D} \sum_\nu \chi^\nu \, \cos \left( \bar \beta^\nu k_0 r_{jk} \right)
\end{equation}
and with $\beta = k \bar \beta$, where $\bar \beta$ is the dimensionless scaled propagation parameter
\begin{equation}
\chi^\nu = \left( \bar \beta^\nu + k_0 \bar \beta^\prime \right) \frac{A}{C^\nu} \left[ e_x^\nu \right|^2.
\end{equation}
The parameters are evaluated numerically for each mode $\nu$.

\end{document}